# Revolutions as Structural Breaks: The Long-Term Economic and Institutional Consequences of the 1979 Iranian Revolution[1]

Nuno Garoupa          Rok Spruk

### Abstract

This paper examines whether major political institutional disruptions produce temporary shocks or structural breaks in long-term development. Using the 1979 Iranian Revolution as a natural experiment, we apply the synthetic control method to estimate its causal effect on economic growth and institutional quality. Drawing on a panel of 66 countries from 1950 to 2015, we construct counterfactual trajectories for Iran in the absence of revolutionary change. Our results show a persistent and statistically significant divergence in per capita GDP, institutional quality, and legal constraints on executive power. We perform in-space and in-time placebo tests to rule out confounding events, such as the Iran-Iraq War and international sanctions, and propose confidence interval estimation to address uncertainty in treatment effects. The findings identify the Iranian Revolution as a structural institutional rupture, with implications for the classification of institutional change more broadly. We contribute a generalizable empirical framework for distinguishing between temporary and structural institutional shocks in long-run development.

[1] Garoupa: Professor of Law, Antonin Scalia Law School, George Mason University, 3301 Fairfax Drive, Arlington VA, E: ngaroup@gmu.edu. Spruk: Assistant Professor of Economics, School of Economics and Business, University of Ljubljana, Kardeljeva ploscad 17, SI-1000 Ljubljana, E: rok.spruk@ef.uni-lj.si. The authors are grateful to Thomas Emery, Martin Gelter, David Gilchrist, Robbert Maseland, Alessandro Melcarne, Giovanni Mellace, Lela Mélon, Juan S. Mora-Sanguinetti, Roland Kirstein, Mitja Kovac, seminar and conference participants at Yale University (Law and Macroeconomics), University of Southern Denmark, Otto von Guericke University of Magdeburg, University of Paris – Nanterre, Católica Porto Business School, and University of Ljubljana for comments, feedback, and initiatives. Constance L. McKinnon and Karyn R. Christensen provided superb research assistance.

# 1    Introduction

Economists have long explored the linkages between political institutions and economic development (Grier and Tullock 1989, North 1990 & 2005, Alesina and Perotti 1994, Weingast 1995, Rodrik et al. 2004, Aidt 2009). Numerous empirical studies have assessed the role of political, legal, and social institutions in shaping the trajectories of economic growth. A key focus in this literature is understanding how changes in political institutions affect long-term growth trajectories (Doucouliagos and Ulubaşoğlu 2008). Building on this stream of research, we analyze the economic and social consequences of a natural experiment: the Iranian Revolution of 1978–79. During this period, Iran's political institutions underwent profound changes, which are likely to have had lasting effects on the country's economic, social, and legal development.

Despite the growing body of research, distinguishing the long-term effects of institutional change remains a challenge. Quasi-natural experiments that allow for credible causal inference on the role of institutions in the process of economic growth and development are rare (Glaeser et al. 2004). In this paper, we refine a method to assess the impact of institutional changes, particularly focusing on major constitutional shifts. Using a sophisticated econometric approach, we construct a synthetic counterfactual to compare Iran's economic performance with and without the shock of the revolution. This particular comparison not only allows for confirming short-term effects of the revolution concentrated around immediate GDP growth fluctuations but also facilitates the study of long-term changes, indicating whether the revolution altered Iran's growth trajectory permanently. Recent advancements in the synthetic control method have improved causal inference, emphasizing design-based approaches (Athey and Imbens 2017), interactive unobserved effects (Xu 2017), matrix completion (Athey et al. 2021), bias correction (Ben-Michael et al. 2021), and penalized estimators (Abadie and L'Hour 2021), among others (Arkhangelsky and Imbens 2024). These methodological advances are crucial for understanding the distinction between temporary and permanent effects of institutional and policy changes.[2]

Our study addresses a critical empirical issue: distinguishing between temporary changes that affect GDP growth without altering long-term trajectories and structural changes that have a lasting impact. This distinction is difficult to pinpoint, as most empirical studies are limited by short observation windows, often only covering a decade or less. As a result, temporary disruptions and structural shifts, especially in political institutions and constitutions, are frequently treated as equivalent, due to the absence of a credible counterfactual. Our approach aims to resolve this challenge by providing an empirical framework that rigorously distinguishes between short-term effects and long-term structural changes.

The contributions of this paper are twofold. First, we apply synthetic control estimation to examine the long-term effects of constitutional changes on economic growth, which has immediate empirical implications for the literature on political economy and development (Grier and Maynard 2016, Geloso and Grier 2022, Gilchrist et al. 2023). Second, we clarify the distinction between temporary and structural changes, particularly by illustrating the differential impacts of constitutional reforms over time, and we subject this distinction to rigorous testing using conventional and more advanced synthetic control estimation techniques.

---

[2] For example, Abadie et al. (2015) used the synthetic control method to examine the economic growth impact of German reunification, while similar methods have been applied to study the effects of armed conflict (Bilgel and Karahasan 2019), economic liberalization (Billmeier and Nannicini 2013), and nuclear disasters (Zheng and Wang 2024) among several others (Gilchrist et al. 2022).

To illustrate our method, we focus on the Iranian Revolution and its subsequent constitutional remaking. The literature on this topic is divided between those who argue that the revolution's effects were short-lived (particularly in light of subsequent liberalization efforts) and those who contend that it represented a true structural break, with long-lasting economic consequences (see, among others, discussion by Walton 1980, Pesaran 1985, Esfahani and Pesaran 2009. Alizadeh and Hakimian 2014). By applying the synthetic control method, we provide evidence that the Iranian Revolution introduced a structural change that had significant, long-term consequences for Iran's growth trajectory, rather than a temporary shock.

It is important to note that our analysis does not attempt to fully disentangle the long-term effects of the revolution from concurrent internal and external policy changes, such as the imposition of international sanctions, the Iran-Iraq War, and subsequent constitutional amendments. These overlapping factors complicate the task of isolating the revolution's impact. To address this, we conduct a series of placebo tests and robustness checks, including the application of synthetic control to specific years in which these concurrent shocks occurred. This approach allows us to assess whether the differences between Iran and its synthetic counterfactual are driven primarily by the revolution or by other contemporaneous factors. Additionally, we account for the interaction of economic effects from the revolution, sanctions, and constitutional changes by constructing empirical confidence intervals for the estimated treatment effect. Our approach thus enables us to capture the intertemporal uncertainty of the revolution's impact. While subsequent turmoil and policy shifts may widen these confidence intervals, they do not undermine the long-term economic significance of the revolution. Our results suggest that the effects of the Iranian Revolution should be interpreted as contingent on time, and we recommend that future studies adopt a confidence interval approach when using synthetic control to capture the evolving impacts of constitutional and policy changes.

The remainder of the paper is structured as follows: Section 2 provides the conceptual framework and historical background of the Iranian Revolution. Section 3 outlines the identification strategy. Section 4 presents the data and empirical methodology, while Section 5 discusses the results. Section 6 concludes.

## 2    Conceptual Framework, Constitutions, and Institutional Theory

### 2.1    Related Literature

In their seminar contribution, Acemoglu et al. (2005) emphasized the critical role of political institutions in shaping economic outcomes, arguing that the fabric of institutions influence growth by constraining economic actors, generating incentives, and determining the organization of production. This particular stream of literature focuses on the distribution of political power, constitutional constraints, and the structure of governance, as key determinants of economic success. Furthermore, Acemoglu and Johnson (2005) distinguished between property rights institutions—those that limit state expropriation and encourage long-term

investment—and contracting institutions, which, although important, are less crucial since individuals often adapt to inefficient contracting environments.[3]

A central theme in this literature is the challenge of institutional reform, with a particular focus on how political institutions can be adjusted to foster long-term economic growth. However, such reforms are often difficult to achieve due to the slow pace of institutional change and the redistribution of political power. While external shocks can catalyze institutional change, sustained economic growth typically requires a combination of factors, including political leadership (Jones and Olken 2005 & 2009) and a window of opportunity (Williamson 2000). The difficulty in assessing the impact of institutional changes is exemplified by natural experiments, such as the differing political regimes in North and South Korea, or the contrasting constitutional developments in various countries, which demonstrate the complex relationship between political institutions and long-term economic outcomes (Spruk et al. 2024).

### 2.2 *Models of institutional changes*

We distinguish between two models of institutional change: evolutionary mutation and punctuated equilibrium. Evolutionary mutation, as articulated by Currie et al. (2016), refers to continuous, incremental changes that cumulatively support long-term growth without triggering major disruptions. In an ideal democratic system, such reforms might include gradual constitutional amendments that reshape political power slowly, without generating abrupt shifts in economic outcomes' trajectories. While these changes are significant over time, they do not result in observable short-term disruptions.

In contrast, Aoki (2000) proposed the punctuated equilibrium model, which posits that long periods of institutional stability are periodically interrupted by significant exogenous shocks, such as economic crises or political upheavals. This model reinforces the resilience of institutions, punctuated by moments of rapid and often profound transformation. Crucially, this framework distinguishes between temporary and structural changes: temporary changes produce short-lived effects on growth and institutional quality, without permanently altering economic development trajectory, whereas structural changes lead to fundamental shifts, redirecting growth paths in significant and long-lasting ways.

Historical events often embody elements of both gradual and disruptive change, making it difficult to classify them within a single model. Our approach aims to identify the dominant features in each case. We categorize an event as a structural change if it leads to a permanent divergence from the pre-existing growth trajectory. By contrast, events with only temporary impacts are classified as temporary changes, while those characterized by slow, incremental evolution are considered gradual changes.

Although theoretical models of institutional change can distinguish between these types of shifts, the empirical literature frequently conflates them due to their similar short-term manifestations. Our methodology provides a rigorous empirical test to differentiate evolutionary mutation, temporary change, and structural change. If an economy experiences gradual change, constitutional shocks should have negligible effects on its long-term growth trajectory. Temporary changes would manifest as significant short-term disruptions without altering the broader trajectory. Structural changes, however, would be evident as a clear divergence between observed outcomes and the synthetic counterfactual, signifying a lasting reorientation of growth trajectories. We illustrate this empirical strategy by examining the Iranian Revolution, which economic policy literature widely acknowledges as a major, potentially structural shift, thus grounding our analysis in a well-documented historical context.

---

[3] These findings seem to contradict the work of La Porta et al. (2008) who suggest, in a different strain of the literature, that both property and contracting institutions (labelled under common and civil law systems) determine successful growth. Other possible explanations, like geography and culture also provide for concurring powerful explanatory variables (Acemoglu and Robinson 2012).

## 2.3 Possible limitations

It is important to recognize that our empirical method does not fully resolve the identification challenges inherent in the study of the relationship between political institutions and long-run development. Our approach estimates synthetic long-run economic and institutional development trajectories under the assumption of no unanticipated shocks. A key challenge lies in the potential spillover effects from trade, conflict, and migration, which can be directly influenced by these shocks. Such spillovers may undermine the validity of our counterfactual assumptions, as separating a treated country from its control group could lead to inaccurate estimates of the counterfactual outcome.

To mitigate these concerns, we exclude countries that experienced the same type of shock in the same year, or any shock that might be directly violating the assumption of (Stable Unit Treatment Value Assignment - SUTVA) as highlighted by Abadie (2021). By excluding simultaneously treated units from the donor pool, we reduce the likelihood of contamination from institutional imitation, ensuring that the control units are not influenced by the same shocks. This particular exclusion also helps to minimize the spillover effects from trade, conflict, migration, and other external factors, as the donor group remains unaffected by the shock experienced by the affected country. However, it is important to interpret the results with caution, particularly for large economies where the potential for spillovers is greater. These concerns are less pronounced for smaller, more open economies, where the impact of external shocks may be more diffused.

## 2.4 The Case of the Iranian Revolution: Constitutional Context

Iran's constitutional history reflects a complex interplay between reformist efforts and traditionalist resistance (Barkhordari 2022). The first constitution of 1906 introduced key modernization elements, such as citizen rights, a bicameral parliament, and a partially elected senate, laying the groundwork for subsequent social and political reforms. Under the Pahlavi dynasty, further efforts aimed at modernizing education, the military, and governance, reducing the influence of religion in politics, and establishing a more representative government, garnered some support (Foroughi 2004).

These modernization efforts significantly influenced the 1979 constitutional revolution. The political thought and institutions underpinning the 1979 constitution emerged as a response to the traditionalist discourse that had opposed modernity and Western liberalism in earlier periods (Barkhordari 2022). For example, the new constitution, influenced by traditionalist Islamic beliefs, emphasized the family as the core societal unit and restricted women's career advancement. However, it was not a complete return to pre-modern religious traditions, nor was it a total break from the political thought that had shaped 20th-century Iranian politics. Notably, there are structural similarities between the 1906 and 1979 constitutions in terms of the political branches of government.

The tension between tradition and modernization profoundly shaped Khomeini's views on political and social Islam. As noted by Barkhordari (2022), Khomeini advocated for the revival of religion and the significant involvement of religious leaders in governance, while simultaneously echoing the arguments of 19th-century liberals who called for equality before the law and constitutional governance. Yet, his constitutional theory diverged from Western liberalism by distinguishing between common law, which had been practiced under previous regimes without obligation, and divine law, which the 1979 constitution codified as binding and obligatory.

The 1979 constitutional revolution, therefore, represents a synthesis of traditional values and responses to earlier constitutional experiments. While modernization politically faltered, as Barkhordari (2022) observes, it nonetheless shaped Iranian politics and the 1979 constitution by refining traditional and religious political discourses. Khomeini's political model of Islamic governance should be seen as an alternative to both Western democracy and constitutional monarchy. It includes elements of the rule of law, such as legal constraints on the executive and legislative branches, and mechanisms for popular ratification through direct elections. At the same time, however, the 1979 constitution retained autocratic elements, echoing features of previous regimes.

*2.5  The Case of the Iranian Revolution: Temporary or Structural Change?*

Current scholarship suggests that the Iranian Revolution aligns more closely with Aoki's punctuated equilibrium model, although some elements of gradual evolution within the Iranian context are also recognized. These gradual changes are largely attributed to factors such as growing social dissatisfaction, the expansion of the financial and social resources of the Islamic clergy during Mohammad Reza Pahlavi's reign, and the suppression of many political movements associated with Marxism, while religious activists were permitted to strengthen their networks. Despite this, a key unresolved issue remains whether the post-revolutionary changes were temporary or structural in nature.

Scholars differ significantly in their assessments of the long-term implications of the revolution on Iran's economic growth trajectories. Much of the debate centers around the role of the oil industry (Salehi-Isfahani 2009a, Bjorvatn et al. 2012 & 2013) and the impact of subsequent liberalization policies (Yavari and Mohseni 2012, Atashbar 2013, Povey 2019). Additionally, the role of Iran's constitutional and institutional framework in shaping economic outcomes has garnered considerable attention. Some argue that Iranian institutions, characterized by rent-seeking and a dysfunctional political economy, hindered growth (Bjorvatn and Selvik 2008, Bjorvatn et al. 2013). Others suggest that both pre-revolution and post-revolution institutions share tendencies toward resource appropriation, implying that the revolution may have resulted in a temporary shock rather than a structural shift.

The long-term effects of the Iranian Revolution remain a subject of significant debate, primarily due to the challenge of constructing a reliable counterfactual. In this context, Farzanegan (2022) made an important contribution by estimating the economic effects of Iran's 1979 regime change and the subsequent Iran-Iraq War. His study estimated the counterfactual trajectory of per capita GDP from 1978 to 1988, assuming the absence of revolutionary upheaval and conflict. By leveraging the growth patterns of Middle Eastern, North African, and oil-producing countries that did not experience similar disruptions, Farzanegan (2022) estimated an average annual per capita income loss of approximately 3,150 USD, or about 40 percent.

Building on this foundational work, our study introduces three key innovations in evaluating the Iranian Revolution's economic impact. First, to assess long-term growth effects, we extend the pre-revolutionary period to 1950–1978 and the post-revolutionary period to 2016. This extended time frame provides a more comprehensive training and validation period, allowing for a more robust comparison of post-revolutionary outcomes with pre-revolutionary benchmarks. Second, to avoid overfitting the outcome trajectory, we progressively expand the donor pool to include countries that did not experience significant internal conflict or political instability during the pre-revolutionary period. This adjustment, informed by data from Brecke (1999) and the Uppsala Conflict Data Program, helps ensure the validity of the SUTVA (Stable Unit Treatment Value Assignment) assumption by excluding countries affected by instability, such as those involved in the Arab Spring protests of 2010, which could otherwise compromise the internal validity of the treatment effect. Third, in addition to evaluating the economic effects of the revolution, we explore its impact on institutional quality, thus contributing to the "beyond GDP" debate (Jones and Klenow 2016). This broader analysis introduces a novel empirical dimension, offering new insights into the revolution's effects that have not been addressed in previous studies.

## 3 Identification Strategy

We observe a finite set of countries $i = 1,2,\ldots J+1$ over the finite time horizon $t = 1,2,\ldots T$. Suppose Iran is exposed to the revolution that starts in the period $T_0 < T$ that lasts without interruption in the full post-treatment period $T_0 + 1, \ldots T$ while the remaining $J$ countries are the potential control units not exposed to the revolution. Our goal is to examine the contribution of the Iranian revolution to the long-term growth and development of Iran. The key variable of interest is the per capita GDP denoted by $Y_{i,t}$. Therefore, our aim is to estimate the effect of the Iranian revolution on the economic growth trajectory of Iran by seeking its counterfactual realization, denoted by $Y_{i,t}^N$. Suppose the growth effect of the revolution is captured by the vector of ex-ante unknown parameters $\{\lambda_{1,T_0+1}, \ldots \lambda_{1,T}\}$ for each period after the revolution where the long-run impact is captured by $\lambda_{1,t} = Y_{i,t} - Y_{i,t}^N$ for each $t > T_0$. Assume that the shock posited by the revolution can be best described by a simple binary switching variable $D \in \{0,1\}$ that takes the value of 1 without interruption in the full post-revolution period $T_0 + 1$. The potential outcome is given by the following factor model:

$$Y_{i,t}^{D \in \{0,1\}} = \begin{cases} Y_{i,t}^N = \phi_t + Z_i \theta_t + \pi_t \mu_i + \varepsilon_{i,t} \\ Y_{i,t} = \lambda_{i,t} + Y_{i,t}^N \end{cases}$$

where $\phi_t$ represents the unobserved time-varying technology shocks common to all countries, $Z_i$ is $(1 \times r)$ vector of observed economic growth covariates, $\theta_t$ is an $(r \times 1)$ vector of unknown parameters, $\pi_t$ is an $(1 \times F)$ vector of unknown common factors, and $\mu_i$ is an $(F \times 1)$ vector of unknown factor loadings. The transitory growth shocks are denoted by $\varepsilon_{i,t}$, where $\varepsilon_{i,t} \sim i.i.d$ is assumed so that shock transitivity restriction $\mathrm{E}\left(\varepsilon_{i,t} | D(1, T_0)\right) = E(\varepsilon_{i,t}) = 0$ holds. The key parameter of interest is $\pi_t \mu_i$ which allows us to capture the time-varying heterogeneous response of economic growth trajectory to the revolution that takes place at $T_0$. To impute the counterfactual scenario of the revolution, the per capita GDP in the hypothetical absence of the revolution, $Y_{i,t}^N$, is not observed to the econometrician and can only be estimated to project a valid representation of the counterfactual. Additional details behind the identification strategy are reported in the supplementary appendix.

In our analysis, the synthetic control method is used to estimate the counterfactual development of Iran's economy in the absence of the 1979 revolution. To construct this counterfactual, we create a weighted combination of countries in the donor pool that share similar pre-revolutionary economic characteristics with Iran. These weights are derived by minimizing the difference between Iran's pre-revolutionary real GDP per capita and the weighted average of the same variable in the donor pool countries. Once the synthetic control is constructed, we compare the actual post-revolutionary growth trajectory of Iran to the synthetic counterfactual, which represents the expected trajectory had the revolution not occurred. This allows us to isolate the long-term effects of the revolution on Iran's economic growth. Importantly, the synthetic control method assumes that the pre-revolutionary period provides an appropriate benchmark for Iran's potential growth path, and the selected donor pool is free from major shocks that could confound the analysis. We also employ a variety of robustness checks to ensure that our counterfactual is not driven by overfitting or external shocks

## 4    Data

### 4.1    *Outcome variables*

To capture and estimate the long-term economic effect of the Iranian revolution, our dependent variable the real GDP per capita adjusted for purchasing power parities (PPP) based on Geary-Khamis conversion into international dollars at 2005 constant prices. The data on per capita GDP is from Bolt and Van Zanden (2014) and is based on the First Update of the Maddison (2007) worldwide GDP per capita database. Moreover, our set of dependent variables reflecting the institutional outcomes comprises a series of established indicators that capture the quality of governance and institutions. In particular, six indicators from updated Varieties of Democracy Dataset (Coppedge et. al. 2022) are considered: (i) women's access to justice, (ii) the degree of economic and political clientelism, (iii) judicial corruption, (iv) judicial constraints on the executive powerholders, (v) equality before the law, and (vi) freedom of expression and alternative information. These indices are designated on the ordinal scale where higher values represent a greater degree of institutional inclusivity, a more egalitarian power distribution, and more broad-based access to collective action to challenge the power of the political elites. Despite the obvious caveats surrounding the aggregation of indices into a singular variable, these variables provide a plausible representation of the institutional fabric and allow us to explore the effects of Iranian revolution on the various paths of institutional development in greater detail beyond the short-term horizon.

### 4.2    *Covariates*

The set of covariates employed in synthetically matching Iran's pre-revolution growth and development trajectory with the rest of the world consists of an expansive set of pre-revolution GDP per capita dynamics and a series of auxiliary covariates. These include demographic, physical geographic, and institutional quality covariates. The set of demographic covariates consists of the fertility rate (number of children born per woman), mortality rate (i.e., per thousand residents), life expectancy at birth, population size and population density. These demographic covariates come from historical United Nations demographic yearbooks and Klein Goldewijk et al. (2017) while the data on population density comes from International Data Base of U.S. Census Bureau. These variables capture the variation in long-term economic growth shaped by the demographic structure to parse out a suitable representation of the synthetic counterfactual for Iran having similar demographic characteristics as the actual Iran prior to the revolution.

The battery of physical geographic covariates consists of the time-invariant characteristics and includes latitude and longitude coordinates, soil quality, size of land area, indicators tropical and desert zones, distance to coast and fraction of area within 100 km of coastline, indicator variables for being an island, landlocked as well as the measure of terrain ruggedness. The data on physical geographic covariates comes from Nunn and Puga (2012). The auxiliary physical geographic characteristics allow for the capture of pre-determined exogenous growth characteristics that are traditionally absorbed by the unobserved heterogeneity bias. These covariates allow us to approximate the synthetic counterfactual trajectory of Iran's economic growth with pre-determined characteristics that are as similar as possible to the ones for the real Iran.

The set of institutional quality characteristics includes variables capturing the type and quality of political institutions. Two specific variables are used for this particular purpose. First, the Polity2 variable (Marshall and Gurr 2020) captures the degree of democracy as a political economy determinant of economic growth. Second, the indicator variable for federation denotes whether a country is a federation (Persson and Tabellini 2005). Third, the indicator variable civil law denotes whether a country in the sample belongs to the civil-law jurisdiction (La Porta et. al. 1998). And fourth, we also construct a dichotomous variable indicating whether the respective country in the sample belongs to the Non-Aligned Movement.[4] For the sake of brevity, we also include an indicator variable denoting whether a country in the sample is in Europe. Taken altogether, these variables roughly capture the type of political regime, degree of democracy as well as legal origin to construct a plausible representation of Iran's economic growth trajectory that is driven by the role of political institutions.

### 4.3  *Samples*

We employ two large samples to estimate the long-term effect of Iranian revolution. Our full sample is used to evaluate the long-term economic growth effect of the revolution. It consists of Iran and 66 independent countries[5] for the period 1950-2016 which yields a strongly balanced panel of 3,886 observations. The first treatment sample consists of Iran given the focus on a single-treatment synthetic control analysis. The control sample consists of those countries where no major armed conflict or similar revolution took place. We rely on the coded armed conflicts by Brecke (1999) to identify country-level episodes of major armed conflict and extend the series until 2016 combining these episodes with the updated coded conflicts from *Uppsala Conflict Data Program*. Our second treatment sample is used to tackle the long-term institutional quality effects of the revolution. It is a more diverse donor pool that consists of 140 independent polities[6] for the period 1955-2021 that corresponds to country-year observation pairs from Coppedge et al. (2020) and Pemstein et al. (2020). This yields a strongly balanced panel of 9,306 country-year observation pairs.

---

[4] The Non-Aligned Movement is a group of 120 developing countries that are not formally aligned with or against any major power bloc. The movement was formally established in 1950s to avoid the polarized world of the Cold War period between the communist states and free-market democracies. Based on the Bandung Conference in 1955, the movement was established in 1961 in Belgrade, former Yugoslavia, through the initiative of several prime ministers such as Jawaharal Nehru (India), Kwame Nkrumah (Ghana), Sukarno (Indonesia), Gamal Abdel Nasser (Egypt) and Josip Broz Tito (Yugoslavia). The movement has been traditionally associated with third-way politics opposing many initiatives of US in foreign policy and support many policies consistent with the idea of democratic socialism. Iran formally joined the movement in the 1970s.

[5] Albania, Australia, Austria, Belgium, Botswana, Bulgaria, Canada, Cape Verde, Costa Rica, Côte d'Ivoire, Denmark, Estonia, Finland, France, Germany, Greece, Honduras, Hong Kong, Iceland, Ireland, Italy, Japan, Luxembourg, Madagascar, Malta, Mauritius, Mongolia, Namibia, New Zealand, Norway, Poland, Portugal, Puerto Rico, Saudi Arabia, Senegal, Singapore, Slovakia, Slovenia, South Africa, South Korea, Spain, Sweden, Switzerland, Thailand, Trinidad and Tobago, Tunisia, Ukraine, United Kingdom, United States, Uruguay, Venezuela, Zambia, and Zimbabwe.

Table 1 presents the pre-revolution GDP per capita dynamics and auxiliary covariate balance between Iran and its synthetic control group. The comparison of Iran with its synthetic peer indicates a reasonable balance of covariates and outcomes prior to the revolution. Panel A depicts pre-revolutionary GDP per capita balances and unveils a rather strong similarity of Iran with its synthetic control group. Notice that the level of real GDP per capita in various benchmark years preceding the revolution shows a very strong similarity between the actual Iran and its synthetic counterpart. This finding suggests that the synthetic control method provides a rather excellent fit of GDP per capita dynamics between Iran and its synthetic counterpart where no major turmoil, revolution, or armed conflict occurred.

**Table 1**: Pre-revolution covariate balance

|  | Real Iran | Synthetic Iran |
|---|---|---|
| *Panel A: Pre-Revolution GDP Per Capita* | | |
| Real GDP Per Capita in 1950 | 2568 | 2003 |
| Real GDP Per Capita in 1960 | 3198 | 3089 |
| Real GDP Per Capita in 1970 | 6173 | 6272 |
| Real GDP Per Capita in 1975 | 8640 | 8629 |
| *Panel B: Demographic Covariates* | | |
| Fertility rate | 6.59 | 4.28 |
| Mortality rate | 234.64 | 108.66 |
| Life expectancy | 46.58 | 57.98 |
| Population size (log) | 10.05 | 9.87 |
| Population density (log) | 2.65 | 5.06 |
| *Panel C: Physical geographic covariates* | | |
| Latitude | 32.56 | 30.31 |
| Longitude | 54.31 | 99.38 |
| Soil quality | 21.82 | 37.03 |
| Land area km2 (log) | 14.31 | 11.71 |
| Desert | 32.56 | 30.31 |
| Tropical | 54.31 | 99.38 |
| Distance to coast | 21.82 | 37.03 |

---

[6] Afghanistan, Albania, Algeria, Angola, Argentina, Australia, Austria, Barbados, Belgium, Benin, Bhutan, Bolivia, Botswana, Brazil, Bulgaria, Burkina Faso, Burundi, Cambodia, Canada, Cape Verde, Chad, Chile, China, Colombia, Comoros, Costa Rica, Cuba, Cyprus, Czech Republic, Congo Dem. Rep., Congo Rep., Cote d'Ivoire, Denmark, Djibouti, Dominican Republic, Ecuador, Egypt, El Salvador, Equatorial Guinea, Eswatini, Ethiopia, Fiji, Finland, France, Gabon, Germany, Ghana, Greece, Guatemala, Guinea, Guinea-Bissau, Guyana, Haiti, Honduras, Hungary, Iceland, India, Indonesia, Iran, Iraq, Ireland, Israel, Italy, Jamaica, Japan, Jordan, Kenya, Kuwait, Laos, Lebanon, Lesotho, Liberia, Luxembourg, Madagascar, Malawi, Malaysia, Maldives, Malta, Mauritania, Mauritius, Mexico, Mongolia, Morocco, Mozambique, Myanmar, Nepal, Netherlands, New Zealand, Nicaragua, Niger, Nigeria, Norway, Oman, Pakistan, Panama, Papua New Guinea, Paraguay, Peru, Philippines, Poland, Portugal, Qatar, Romania, Russia, Rwanda, Sao Tome and Principe, Saudi Arabia, Senegal, Seychelles, Sierra Leone, Singapore, Solomon Islands, Somalia, South Africa, South Korea, Spain, Sri Lanka, Sudan, Suriname, Sweden, Switzerland, Syria, Taiwan, Tanzania, Thailand, The Gambia, Togo, Trinidad and Tobago, Tunisia, Türkiye, Uganda, United Kingdom, United States, Uruguay, Vanuatu, Venezuela, Vietnam, Yemen, Yugoslavia, Zambia, and Zimbabwe.

|  |  |  |
| --- | --- | --- |
| Fraction of area within 100 km coastline | 0.54 | 0.82 |
| Landlocked | 0 | 0.01 |
| Island | 0.46 | 0.08 |
| Terrain ruggedness | 2.45 | 2.02 |
| *Panel D: Institutional quality covariates* | | |
| Federation | 0 | 0.02 |
| Civil law | 1 | 0.91 |
| Polity2 | -8.61 | 0.95 |
| Europe | 0 | 0 |
| Non-aligned movement | 1 | 0.21 |

### 4.4 Descriptive analysis

Figure 1 presents the trajectories of per capita GDP for Iran, electoral and liberal democracy and the rest of world. Notwithstanding further empirical analysis, the pattern reveals a clear indication of accelerated per capita GDP convergence of Iran with the rest of the world in the pre-revolution period as well as a marked and substantial divergence in the post-revolution period up to the present day. From the comparison, two features become apparent. First, Iran's economic growth trajectory is characterized by a notable acceleration up to the revolution period which appears to move substantially closer to the sample mean. And second, the trajectory appears to be derailed in the post-revolutionary period and despite some improvement after the end of Iran-Iraq war, it loses both pace and parity with the rest of the world. For the trajectories of electoral and liberal democracy indices, once again, the descriptive patterns reveal a straightforward indication that the trajectory of political and institutional development fails any convergence in the post-revolution period up to the present day.

**Figure 1**: Economic growth trajectories of Iran and the rest of the world, 1950-2016

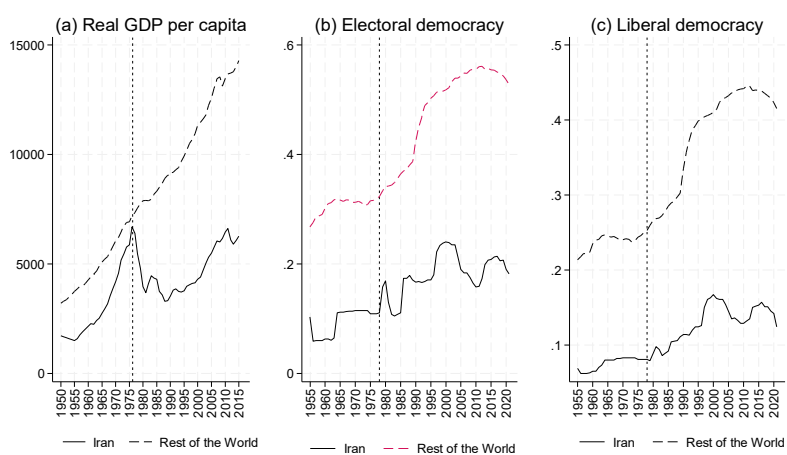

## 5 Results

### 5.1 Baseline results

Table 2 presents the composition of synthetic control groups across the baseline and restricted specifications along with the root mean square prediction error (RMSPE). The size of RMSPE appears to be very low. Leveraging the RMSPE against its null version (Adhikari and Alm 2016), the estimated margin of error is less than 1 percent of the pre-treatment window margin which confirms a reasonably good fit between Iran and its synthetic control group prior to the revolution. In the baseline specification, Iran's pre-1976 economic growth trajectory is best reproduced as a convex combination of the implicit growth and development characteristics of South Korea (48 percent), Japan (30 percent), Cape Verde (14 percent), South Africa (6 percent), Canada (2 percent) and Saudi Arabia (2 percent), respectively. The composition of the synthetic control groups does not seem to shift markedly across the restricted specifications. Across the full set of replications, the synthetic control groups are dominated by South Korea and Japan followed by countries such as Mongolia, Tunisia, Honduras, New Zealand, and Saudi Arabia. The composition of control groups indicates that prior to the revolution, Iran's growth trajectory appears to be very similar to that of rapidly growing East Asian economies and a few others with similar exogenous geographic characteristics such as South Africa, Tunisia, and Cape Verde. Persistent deterioration of the growth trajectory after the revolution indicates that without the revolution, Iran would most likely be a developed economy down to the present day whilst the contemporary per capita GDP is about one half below the possibility frontier (due to the bad institutional design undermining a high potential for rapid economic growth and modernization).

**Table 2**: Composition of Iran's synthetic control groups

|  | Full donor pool | w/o pre-1955 observations | w/o pre-1960 observations | w/o pre-1965 observations |
|---|---|---|---|---|
| Average treatment effect | 0.491 | 0.541 | 0.477 | 0.507 |
| (95% confidence bounds) | [0.44, 0.54] | [0.48, 0.59] | [0.43, 0.53] | [0.43, 0.53] |
| End-of-sample treatment effect | 0.499 | 0.587 | 0.486 | 0.510 |
| RMSPE | 216.78 | 121.49 | 133.12 | 135.50 |
| Albania | 0 | 0 | 0 | 0 |
| Australia | 0 | 0 | 0 | 0 |
| Austria | 0 | 0 | 0 | 0 |
| Belgium | 0 | 0 | 0 | 0 |
| Botswana | 0 | 0 | 0 | 0 |
| Bulgaria | 0 | 0 | 0 | 0 |
| Canada | 0.02 | 0 | 0 | 0 |
| Cape Verde | 0.14 | 0 | 0.08 | 0 |
| Costa Rica | 0 | 0 | 0 | 0 |
| Côte d'Ivoire | 0 | 0 | 0 | 0 |
| Denmark | 0 | 0 | 0 | 0 |
| Estonia | 0 | 0 | 0 | 0 |
| Finland | 0 | 0 | 0 | 0 |
| France | 0 | 0 | 0 | 0 |
| Germany | 0 | 0 | 0 | 0 |
| Greece | 0 | 0 | 0 | 0 |
| Honduras | 0 | 0 | 0.01 | 0 |
| Hong Kong | 0 | 0 | 0 | 0 |
| Iceland | 0 | 0 | 0 | 0 |
| Ireland | 0 | 0 | 0 | 0 |
| Italy | 0 | 0 | 0 | 0 |
| Japan | 0.30 | 0.32 | 0.32 | 0.34 |
| Luxembourg | 0 | 0 | 0 | 0 |
| Madagascar | 0 | 0 | 0 | 0.11 |
| Malta | 0 | 0 | 0 | 0 |
| Mauritius | 0 | 0 | 0 | 0 |
| Mongolia | 0 | 0.03 | 0 | 0.09 |
| Namibia | 0 | 0 | 0 | 0 |
| New Zealand | 0 | 0 | 0.01 | 0 |
| Norway | 0 | 0 | 0 | 0 |
| Poland | 0 | 0 | 0 | 0 |
| Portugal | 0 | 0 | 0 | 0 |
| Puerto Rico | 0 | 0 | 0 | 0 |
| Saudi Arabia | 0.02 | 0.02 | 0.02 | 0.03 |
| Senegal | 0 | 0 | 0 | 0 |
| Singapore | 0 | 0 | 0 | 0 |
| Slovakia | 0 | 0 | 0 | 0 |
| Slovenia | 0 | 0 | 0 | 0 |
| South Africa | 0.06 | 0 | 0 | 0 |
| South Korea | 0.48 | 0.36 | 0.48 | 0.42 |
| Spain | 0 | 0 | 0 | 0 |
| Sweden | 0 | 0 | 0 | 0 |
| Switzerland | 0 | 0 | 0 | 0 |
| Thailand | 0 | 0 | 0 | 0 |
| Trinidad and Tobago | 0 | 0 | 0 | 0 |
| Tunisia | 0 | 0 | 0.08 | 0 |
| Ukraine | 0 | 0 | 0 | 0 |
| United Kingdom | 0 | 0 | 0 | 0 |
| United States | 0 | 0 | 0 | 0 |
| Uruguay | 0 | 0 | 0 | 0 |
| Venezuela | 0 | 0 | 0 | 0 |
| Zambia | 0 | 0 | 0 | 0 |
| Zimbabwe | 0 | 0.27 | 0 | 0 |

Figure 3 presents the synthetic control estimates of the long-term economic impact of the Iranian Revolution. The solid line represents the actual economic growth trajectory of Iran, while the dashed line reflects the synthetic counterfactual, representing the hypothetical trajectory of Iran's economy had the revolution not occurred. Our baseline estimate matches the growth trajectory of Iran and its synthetic control group for the full training and validation period from 1950 to 1975. The results indicate a substantial and persistent economic loss stemming from the revolution, as evidenced by the clear divergence between the actual and synthetic growth paths. In the counterfactual scenario, Iran's economic trajectory would have likely experienced substantial improvements in the absence of the revolution.

The gap in per capita GDP between Iran and its synthetic control group is both immediate and persistent, showing no indication of convergence over time. Specifically, the end-of-sample per capita GDP ratio between Iran and its synthetic control group is 0.49, with a 95% confidence interval ranging from 0.44 to 0.54. This suggests that, as of the end of the sample period, Iran's per capita GDP is between 46% and 54% lower compared to its synthetic peer—a country that did not undergo the same major disruptions, such as armed conflict. This comparison strongly suggests that the Iranian Revolution was not a temporary shock, but rather the onset of a structural breakdown in economic growth. The real Iran's economic performance fails to recover to pre-revolutionary levels, as implied by the trajectory of its synthetic control group.

In the hypothetical absence of the revolution, Iran's per capita GDP by the end of the sample would have reached approximately 31,000 USD, placing its economic performance in line with countries such as Slovenia or Spain. In contrast, the actual per capita GDP of Iran at that time aligns more closely with that of Bulgaria and Serbia. We replicate our baseline estimates across various sub-periods, starting with pre-treatment years of 1955, 1960, and 1965. In all cases, the end-of-sample per capita GDP ratio between Iran and its synthetic peer ranges from 0.48 to 0.59, indicating that Iran's contemporaneous per capita GDP is between 41% and 52% lower than that of the synthetic control group, which represents the growth path in the absence of the revolution.

It is important to note that while the per capita GDP gap between Iran and its synthetic control group narrows following economic liberalization and structural reforms in the 1990s, this narrowing is insufficient to reverse the long-term derailment of Iran's growth trajectory caused by the revolution. Figure 4 tracks the evolution of the per capita GDP ratio between Iran and its synthetic peer throughout the post-revolutionary period, alongside the 95% confidence interval (Firpo and Possebom 2018). This comparison indicates some reduction in the GDP gap beginning with the economic reforms of the 1990s, but there is no evidence to suggest that these reforms have been powerful enough to reverse the structural impact of the revolution. Moreover, after 2010, the gap widens once again, suggesting a further deepening of the economic penalty initiated by the revolution.

**Figure 3**: Long-term economic growth effect of Iranian revolution, 1950-2016

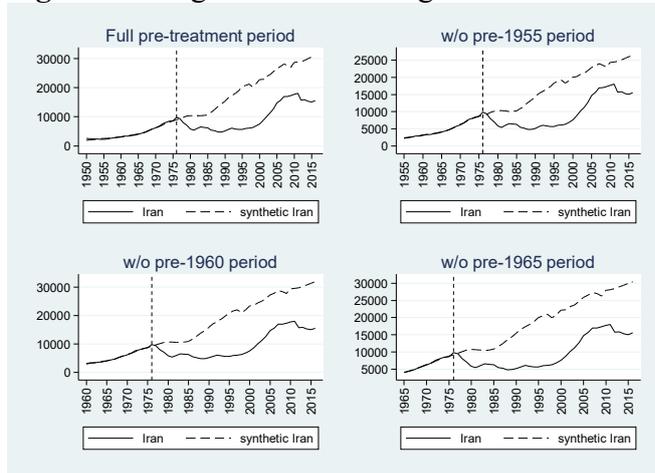

**Figure 4**: Aggregate uncertainty in the long-term economic growth effect of Iranian revolution

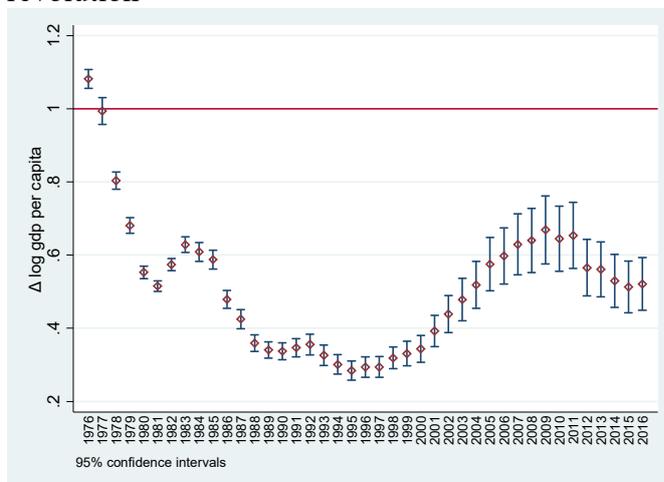

*5.1.2 In-space placebo analysis*

To assess the robustness of the economic growth impact attributed to the Iranian Revolution, we ask a fundamental question: Could our results be driven purely by chance? Specifically, how likely would we be to obtain an estimate of such a large magnitude if we selected a country at random instead of Iran and assign it the shock of the revolution? To address this question, we conduct a series of placebo analyses. In these analyses, we apply the synthetic control method to countries in our donor pool that did not experience a revolution during the study period. The intuition behind this placebo test is straightforward: since the countries in the donor pool did not undergo the revolutionary turmoil that Iran experienced, if the placebo analysis produces per capita GDP gaps of a magnitude similar to that of Iran's, then the observed negative growth impact of the revolution would be less meaningful. Conversely, if the placebo tests reveal that the per capita GDP gap for Iran is significantly larger than those for countries without revolutionary upheaval, it strengthens the case for a substantial and negative impact of the Iranian Revolution on Iran's long-term economic growth.

In this context, we follow the inference procedures proposed by Abadie et al. (2010) based on treatment permutation methods. Specifically, we apply the synthetic control method to each untreated country in the donor pool, estimating the effect of the Iranian Revolution for each. For single-treatment case studies, the permutation distribution is derived by reassigning the underlying political shock to countries in the donor pool, thus generating a distribution of placebo effects from each iteration. By pooling the estimated effects for Iran along with those for the other countries, we construct the permutation distribution to assess the statistical significance of our estimate. Since countries in the donor pool may not closely follow Iran's pre-revolutionary GDP trajectory, we adopt the decision rule outlined by Abadie et al. (2010) and Abadie (2021). This involves comparing the post-revolution fit to the pre-revolution fit, relative to the pre-revolution fit, using this test statistic to measure the quality of the fit in the post-revolutionary period. Additionally, to ensure that placebo effects are not inflated by an unusually poor fit, we discard placebo simulations where the root mean square prediction error (RMSPE) is multiple times larger than the RMSPE for Iran. The p-values for the statistical inference are derived from the permutation distribution (Galiani and Quistorff 2017, Firpo and Possebom 2018).

Figure 5 presents the results from this in-space placebo analysis. More specifically, it represents the distribution of p-values on the null hypothesis of zero economic growth impact of the Iranian revolution. The evidence largely confirms the significance of the growth losses emanating from the revolution. Notice that the p-values indicate the proportion of countries from the placebo analysis that have at least as large RMPSE as Iran. If the proportions were high, much of the post-1975 per capita GDP gap would be easily attributed to chance or lack of fit. By contrast, our analysis reveals that the proportion of countries with RMSPE at least as large as that of Iran is well below a 10 percent threshold until the late 1990s whilst within the 20 percent threshold afterwards. The pattern of p-values confirms a strong negative effect of the revolution in terms of the lost economic growth relative to the counterfactual scenario. The effect appears to have been mildly moderated by the economic liberalization in the 1990s but not to the degree that would have rendered the nature of the Iranian revolution temporary in the long run. To address this potential issue, we perform several placebo simulations, progressively excluding countries whose pre-revolution RMSPE is large relative to Iran's. In particular, we exclude countries with an RMSPE four or five times larger than Iran's, and the results continue to show a significant and easily detectable negative per capita GDP gap for Iran. Moreover, we further refine our placebo analysis by excluding countries with RMSPE more than twice the size of Iran's pre-revolution RMSPE, ensuring that we focus only on countries that closely track Iran's pre-revolution GDP trajectory between 1950 and 1975. In these specifications, the estimated per capita GDP gap for Iran remains among the largest among all comparisons. These results confirm that the unusually large per capita GDP gap for Iran is not driven by issues with the fit, but rather reflects a genuine, significant negative growth impact of the Iranian Revolution.

**Figure 5**: Inference on the economic growth impact of Iranian revolution

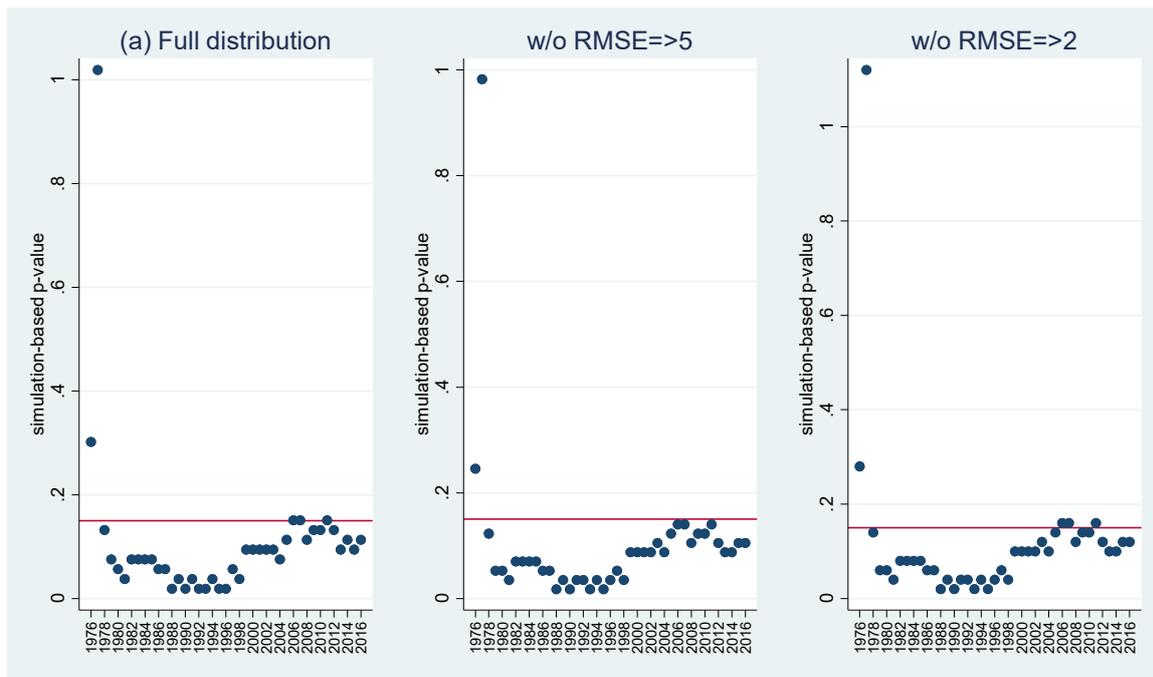

To clarify the methodology used in the in-space placebo analysis, the purpose of the analysis to assess the statistical significance of the Iranian Revolution's economic impact by comparing Iran's actual per capita GDP gap with those generated by the synthetic control method for countries in the donor pool that did not experience the revolution. To this end, we compute the root mean square prediction error (RMSPE) for each country in both the pre-revolutionary and post-revolutionary periods, and compare the ratio of Iran with the estimated placebos A key expectation is that the RMSPE for Iran in the post-revolutionary period should be the highest among the donor countries, as the revolution represents a structural break in Iran's economic trajectory, and Iran's growth path diverges most significantly from the counterfactual. The approximated p-values on the null hypothesis in Figure 5 reflect the relative uniqueness of the per capita GDP gap, which naturally fluctuates over time for different countries, rather than a consistent increase. The key finding is that Iran's post-revolutionary gap is unusually large compared to those of the placebo countries. Additionally, conventional tests for differences in variances or standard deviations are not applicable to the synthetic control method, as it constructs a counterfactual trajectory by weighing donor countries to best match pre-revolutionary outcomes. The appropriate statistical inference is derived from permutation tests, which compare the actual effect to the distribution of placebo effects, ensuring that we account for the significance of Iran's estimated treatment effect relative to the placebo distribution.

### 5.1.3          *In-time placebo analysis*

The evidence presented thus far stresses a significant negative effect of the Iranian Revolution on Iran's long-term economic growth, with a discernible persistence over time. The magnitude of this negative impact appears to be unique to Iran, as it does not manifest similarly when the synthetic control method is applied to other countries in the donor pool. While the in-space placebo analysis strengthens the internal validity of our findings, it remains critical to further test the robustness of the results by examining whether the synthetic control method also produces similarly large effects when applied to dates that do not coincide with the Iranian Revolution. Several potential threats to the internal validity of the estimated effect warrant attention. For instance, the post-revolutionary per capita GDP gap could reflect the effects of external factors, such as international sanctions imposed on Iran in the early 1980s, the end of the Iran-Iraq War in 1988, or subsequent constitutional changes. Without addressing these additional events, the plausibility of the observed per capita GDP gap could be called into question.

To mitigate these concerns, and following established practices in the literature (Heckman and Hotz 1989, Abadie et al. 2015), we apply the synthetic control estimator to a series of alternative years to test whether similarly large effects on Iran's growth trajectory can be attributed to events other than the revolution. We perform the Zivot and Andrews (1992) structural break test, which allows us to identify the year with the most significant structural break in the per capita GDP series. This test identifies 1981 as the year with the largest structural break in the growth trajectory, coinciding with the onset of the Iran-Iraq War. This year is a logical candidate for our in-time placebo analysis, as it represents a major external shock that could potentially confound the effects of the revolution.

Figure 6 presents the results of the in-time placebo analysis, where we assign the year of the revolution to the year 1981. Importantly, this test yields no evidence of similarly large effects on the growth trajectory. The pre-treatment fit quality is significantly worse than in the baseline analysis, confirming that the synthetic control for Iran provides a much better match to Iran's growth trajectory prior to the revolution. The break between the actual and synthetic growth trajectories occurs precisely in 1979, aligning with the revolution, rather than in 1981, reinforcing the idea that the observed economic downturn is attributable to the revolution itself rather than to the onset of the Iraq-Iran War. Furthermore, post-1981, the growth trajectories of Iran and its synthetic control group closely align, empirically ruling out the possibility that the estimated post-revolutionary growth effect is driven by the war or other later shocks.

The root mean square prediction error (RMSPE) for the in-time placebo analysis is 732.72, more than three times larger than the baseline estimate. This substantial difference in RMSPE further supports the conclusion that the estimated long-term negative growth effect is robust and specific to the Iranian Revolution. To further test the robustness of our results, we perform additional placebo analyses using other potential break years, including: (i) the end of the Iran-Iraq War in 1988, (ii) the constitutional amendment in 1989, (iii) the onset of economic sanctions in 1981, and (iv) the implementation of the general economic reforms outlined in Principle 44 of the Iranian Constitution in 2006. For each of these placebo years, we find no evidence of a conflicting effect with the baseline estimated gap. Therefore, our in-time placebo analysis confirms that the estimated per capita GDP gap post-revolution cannot be attributed to the policies following the war, the constitutional amendments, or the later economic reforms of the 2000s.

**Figure 6**: In-time placebo analysis of the long-term growth effect of Iranian revolution

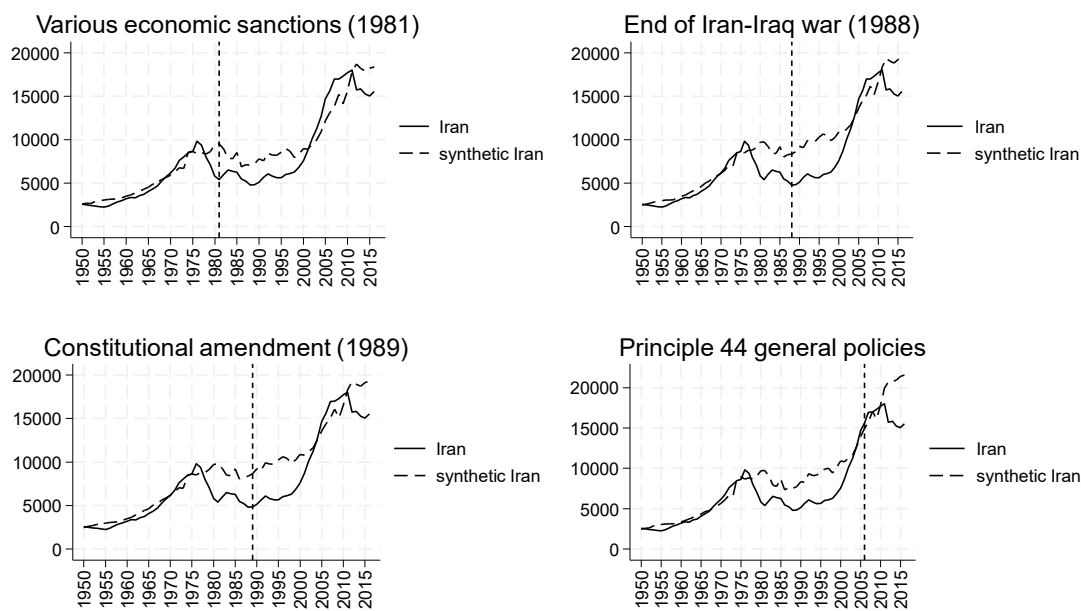

### 5.2 Institutional quality effects of Iranian revolution

Figure 7 presents the estimated long-term effects of the Iranian Revolution on a range of institutional quality outcomes. Using the synthetic control method outlined in Section 3, we estimate the counterfactual trajectories of institutional performance in the absence of the 1979 revolution and constitution. The pre-revolutionary fit between Iran and its synthetic counterpart is remarkably strong across all institutional dimensions, with root mean square prediction errors (RMSPE) within 1% of the benchmark null (Adhikari and Alm 2016), highlighting the internal validity of the synthetic control estimator in reproducing pre-revolutionary institutional dynamics and the robustness of the baseline estimates.

Our findings indicate a marked and persistent deterioration in institutional quality following the adoption of the 1979 constitution. One of the most salient outcomes is the collapse of women's access to justice, which, in contrast to steady improvement during the pre-revolution period, exhibits a sharp and sustained decline thereafter. The post-revolution gap between Iran and its synthetic control is statistically significant at the 1% level and shows no signs of narrowing over time, indicating a structural and likely irreversible setback in institutionalized access to justice for women. Our results also indicate a sharp and unequivocal escalation of political and economic clientelism following the revolution. The average post-revolution gap in clientelism indicators relative to synthetic control is substantial and statistically significant at the 1% level, pointing to a widespread entrenchment of patronage-based governance. Additionally, the data reveal a pronounced increase in judicial corruption, a trend that coincides with the growing influence of Islamic jurisprudence in the composition and operation of key legal institutions, such as the Supreme Council.

While Iran's trajectory of judicial corruption prior to the revolution closely mirrors that of its synthetic counterpart, a divergence emerges in the early 1980s that persists to the end of the sample period. The underlying escalation of corruption is closely linked with the systematic erosion of judicial constraints on the executive, which our estimates show to have deteriorated sharply after the implementation of the 1979 constitution. In the absence of the revolution, our counterfactual estimates suggest that Iran's level of judicial constraint would today be comparable to that of Southern European democracies such as Italy or Slovenia. Instead, the observed trajectory has placed Iran's institutional profile closer to that of contemporary authoritarian regimes such as Russia and China, with significant implications for rule of law and accountability.

Beyond these specific outcomes, the results from our analysis also emphasize the revolution's detrimental impact on broader principles of legal equality and freedom of expression. Across both indicators, the null hypothesis of no post-revolutionary gap is rejected at the 1% level. While the actual trajectories show modest gains in the post-2000 period, the synthetic counterfactuals consistently suggest that, in the absence of the 1979 revolution and its accompanying constitutional framework, both legal equality and freedom of information would have advanced considerably. These findings reinforce the interpretation of the Iranian Revolution as a structural institutional rupture, with enduring and broadly regressive effects on the quality of governance and individual rights—effects that have, in turn, contributed to Iran's long-run economic underperformance.

**Figure 7**: Long-term effect of Iranian revolution on institutional quality, 1955-2021

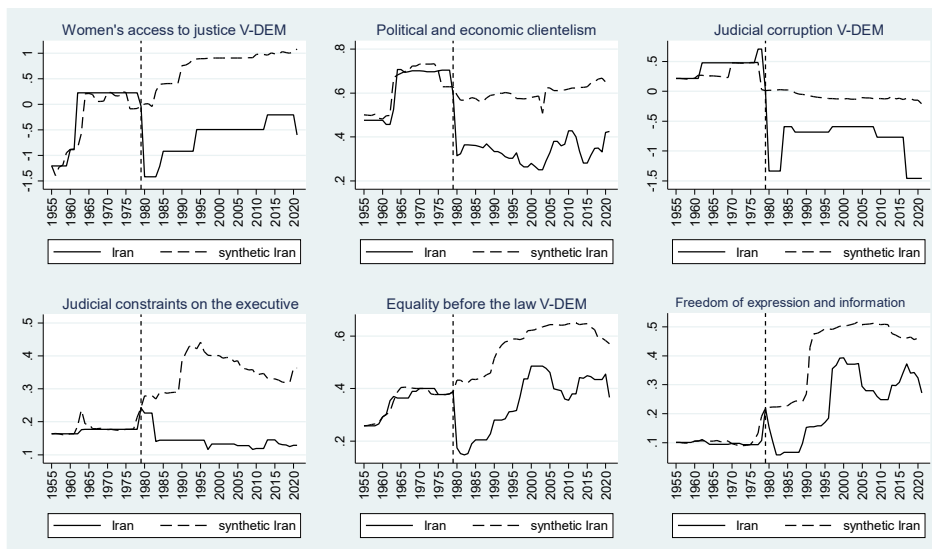

The evidence presented thus far points to a significant deterioration in both institutional quality and structure following the adoption of the 1979 constitution. The underlying deterioration reflects a fundamental reorientation of Iran's institutional framework—marked by a curtailment of civil liberties, diminished access to justice, and a shift away from inclusive, participatory governance toward a more extractive system that served the political objectives of the revolutionary elite. A key question that follows is the extent to which these institutional changes extended beyond political and legal domains to affect economic institutions more broadly. While long-run, cross-national data on economic institutions are limited, we partially address this gap by using the V-Dem Index of State Ownership of the Economy as a proxy for institutional inclusivity in the economic domain. This measure captures the relative dominance of the state in the economy and serves as an inverse indicator of private-sector development and market openness.

Figure 8 presents the estimated effect of the 1979 constitution on the trajectory of state ownership in Iran. The synthetic control group closely replicates Iran's pre-revolution trend, indicating a strong quality of fit and suggesting that the counterfactual trajectory is credible. Post-revolution, the divergence between Iran and its synthetic peer is both substantial and persistent. The point estimates indicate a large and statistically significant increase in state ownership following the revolution, with the estimated gap widening over time and remaining significant at the 1% level across the entire post-treatment period. These results suggest that the revolution significantly intensified the state's role in the economy, crowding out private-sector activity and entrenching state control over key economic assets. Taken together, these findings point to a transformation in Iran's economic institutions that parallels the degradation of political and legal institutions. The shift toward greater state ownership emerges as a key channel through which the structural economic stagnation and long-run per capita GDP losses—documented in earlier sections—have been realized. This evidence supports the interpretation of the 1979 revolution not merely as a political realignment, but as a comprehensive institutional rupture with enduring adverse consequences for economic development.

**Figure 8**: Long-term effect of Iranian revolution on state ownership in the economy, 1955-2021

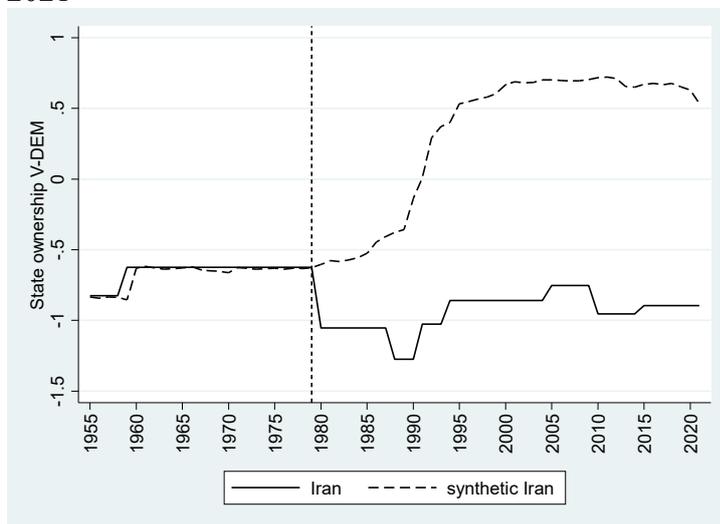

Figure 9 presents the cumulative distribution of non-zero weights assigned to countries in the donor pool used to construct the synthetic control for Iran for the full range of institutional quality outcomes from Figure 7. These weights reflect the relative contribution of each country to the synthetic version of Iran, capturing how frequently and substantively a given donor country is used across model specifications to provide a replica of Iran's institutional quality trajectory prior to the 1979 revolution. The distribution provides insight into the composition and robustness of the synthetic control, indicating which countries most closely resemble Iran in terms of pre-revolution institutional characteristics. Given the relatively low degree of pre-treatment imbalance, we interpret the synthetic control group as offering a credible and policy-relevant counterfactual for institutional development in the absence of revolutionary disruption.

Among the most influential donor countries with non-zero weights and high frequencies, we identify Kenya, Djibouti, Albania, China, Algeria, and Argentina. The cumulative distribution of these weights exhibits a declining pattern reminiscent of Benford's Law, wherein a small number of donor countries contribute most of the explanatory power, while the remainder contribute marginally. To formally assess the distributional properties of these weights, we conduct a two-sample, non-parametric Kolmogorov–Smirnov (K-S) test, comparing the empirical cumulative distribution of the non-zero donor weights to a reference distribution derived from Benford's Law. To evaluate the asymptotic behavior of the empirical distribution as the number of independent and identically distributed observations increases, we also compute the corresponding Glivenko–Cantelli p-value. Under the null hypothesis, the empirical cumulative distribution function (ECDF) of donor weights follows the reference distribution. The test statistic measures the maximum distance between the ECDF and the theoretical cumulative distribution. Our results strongly reject the null hypothesis of distributional equivalence (p-value = 0.000), confirming that the distribution of non-zero donor weights is not an artifact, but instead exhibits systematic structure consistent with Benford law dynamics. This pattern reinforces the interpretive validity of the synthetic control model: a limited but diverse set of countries plausibly approximate the institutional trajectory of Iran in the absence of the revolution, while the declining weight distribution ensures that the counterfactual is not dominated by a single donor country.

**Figure 9**: Benford distribution of cumulative weight density for Iran's synthetic control groups

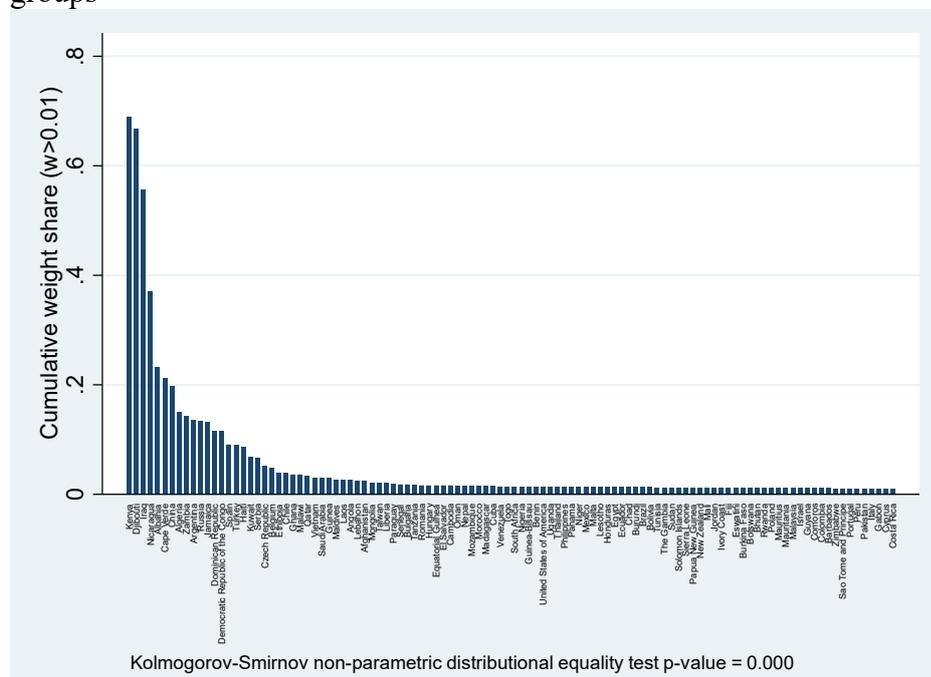

## 6 Conclusion

This article has examined the long-term effects of institutional change on economic growth and institutional development by applying a rigorous synthetic control framework to the case of the 1979 Iranian Revolution. Our contribution lies in estimating the evolution of a counterfactual trajectory—what Iran's economic and institutional path might have looked like in the absence of revolutionary upheaval—using a carefully constructed set of donor countries that share comparable pre-revolution characteristics of economic growth and institutional development. By distinguishing between temporary shocks and structural breaks, we offer a systematic empirical strategy to evaluate whether institutional changes produce short-run fluctuations or permanent deviations in economic growth and institutional outcomes. The Iranian Revolution, long the subject of rich historical and theoretical debate, provides a unique setting for such an analysis. Our findings offer a clear verdict: the revolution represents a structural transformation, not a transitory disturbance.

Our identification strategy relies on two complementary samples and donor pools, drawing on countries that remained free from revolutionary disruption during the 1950–2015 period. By modelling per capita GDP and institutional quality trajectories, we uncover a persistent and statistically significant divergence between Iran and its synthetic counterpart following the revolution. The synthetic control estimates reveal a permanent derailment of both economic and institutional development trajectories. The deterioration of economic growth and institutional development trajectories is robust to a battery of in-space and in-time placebo tests, which confirm that the magnitude and persistence of the observed effects are unique to Iran and not replicable across other countries or time periods. While post-revolutionary developments—such as the Iran–Iraq War, international sanctions, and constitutional amendments—may have shaped the precise evolution of outcomes, these post-revolutionary shocks and changes do not explain away the core finding: the revolution marks a decisive break with Iran's previous development trajectory.

To further address concerns about overlapping policy shocks, we conduct placebo tests by reassigning the revolutionary intervention to a series of false treatment years. The results indicate that the estimated gaps in economic growth and institutional quality do not emerge outside the actual 1979 event. These falsification exercises strengthen the causal interpretation of the revolution's long-term impact. In addition, we advocate for the use of empirical confidence intervals in synthetic control analyses to more effectively quantify uncertainty in long-term growth outcomes, especially in settings where overlapping shocks complicate the identification of treatment effects.

Beyond its substantive focus on Iran, our study introduces a framework for classifying institutional change through counterfactual modeling. By operationalizing synthetic control to track the divergence between observed and hypothetical trajectories, we offer a method for identifying whether a given institutional shift constitutes a gradual adjustment, a temporary disruption, or a permanent structural transformation. While our approach does not explain the political or social mechanisms that determine which institutional changes persist, it provides an empirical basis for future research to explore why certain reforms take root while others dissipate.

Within the broader literature on the Iranian Revolution, our findings align with interpretations that regard the 1979 constitution as a fundamental rupture rather than a reformist detour. The revolution did not merely delay Iran's growth and institutional modernization—it reoriented its trajectory. Despite later economic liberalization and formal amendments, the institutional degradation that followed the revolution remains evident today in diminished judicial independence, increased state control of the economy, curtailed civil liberties, and weaker legal protections. These legacies continue to constrain Iran's long-term growth potential, and more broadly, offer a cautionary example of how deeply institutional shocks can reconfigure the developmental paths of nations.